\documentclass{article}
\usepackage{latexsym}
\usepackage{epsfig}
\begin{document}

\title{Resonance Behavior and Partial Averaging \\
in a Three-Body System with Gravitational Radiation Damping}
\author{Zachary E. Wardell\thanks{e-mail: zew554@mizzou.edu}
\\Department of Physics and Astronomy\\
University of Missouri-Columbia\\Columbia, Missouri 65211, USA}

\maketitle
\begin{abstract}
In a previous investigation, a model of three-body motion was developed 
which included the effects of gravitational radiation reaction. 
The aim was to describe the motion of a relativistic binary pulsar
that is perturbed by a third mass and look for resonances between the
binary and third mass orbits. 
Numerical integration of an equation of relative motion that
approximates the binary gives evidence of such resonances.
These $(m:n)$ resonances are defined for the present purposes by the resonance 
condition, $m\omega=2n\Omega$, where $m$ and $n$ are relatively prime integers 
and $\omega$ and $\Omega$ are the angular frequencies of the binary orbit and 
third mass orbit (around the center-of-mass of the binary), respectively.
The resonance condition
consequently fixes a value for the semimajor axis $a$ of the binary orbit for the
duration of the resonance because of the Kepler relationship $\omega=a^{-3/2}$.
This paper outlines a method of averaging developed by Chicone, Mashhoon, and
Retzloff which renders a nonlinear system that undergoes resonance capture
into a mathematically amenable form. This method is applied to the
present system and one arrives at an analytical solution that
describes the average motion during resonance. Furthermore, prominent features 
of the full nonlinear system, such as the frequency of oscillation and 
antidamping, accord with their analytically derived formulae. 
\\
\\
\textbf{Key words:} celestial mechanics, relativity, gravitational waves
\end{abstract}

\section{INTRODUCTION}

A model was established in a previous paper, `Gravitational Radiation Reaction
and the Three Body Problem' (Wardell 2002), in which the motion of a binary 
system was studied.
This binary system was subject to gravitational radiation damping and the
gravitational influence of a third mass. The equations of motion for the relative
motion of the binary were derived with certain approximations that would
highlight the effects under investigation and also make analysis easier. 
For example, the motion of the three masses is taken to be planar and the
center-of-mass of the binary moves in a fixed circular orbit around
the third mass. Furthermore, the distance of the third mass from the binary's
center-of-mass is taken to be substantially larger than the size of the
relative orbit of the binary. The masses are considered to be point masses. 
After a scaling transformation that makes the variables dimensionless, one 
arrives at the following form of the equation of motion in Cartesian coordinates:
\begin{equation}
\frac{d^{2}r^{i}}{dt^{2}} = -
\frac{r^{i}}{r^{3}}
-\epsilon  K_{ij}r^{j}-\delta R^{i}.
\end{equation}
The variable $r^{i}$ represents the relative orbit, $K_{ij}(t)$ is the tidal 
matrix
that retains the information about the tidal interaction between the third
mass and the binary system, and $R^{i}$ is the radiation reaction term that
expresses the radiation reaction force to desired order after iterative
reduction. To avoid `runaway solutions' that can arise as the result of the
radiation reaction perturbation which involves a fifth time derivative,
one can apply the method of iterative reduction. This gives rise to a
set of equations that is second order and resembles an equation of motion
in Newtonian mechanics (Chicone et al. 2001).
One recovers the differential equation for the appropriate Kepler problem if 
$\epsilon=\delta=0$.

Numerical analysis of the system has revealed that given appropriate 
initial conditions, one
arrives at a result which shows resonance behavior in the relative orbit.
When the graph of the Delaunay variable $L$, where $L=a^{1/2}$ such that
$a$ is the semimajor axis of the osculating ellipse of the relative orbit, 
is plotted versus time one 
sees that the trend of semimajor axis decay can temporarily stop on average at 
a resonance. 
An oscillation occurs around a fixed average value for $L$ for the duration of 
this resonance. This
resonance capture is indicative of an average balance of energy that leaves the
binary system by way of gravitational waves and enters the system because
of the tidal gravitational influence of the third mass.

This paper concerns itself with the behavior of the system when a resonance
occurs; that is, when the resonance condition $m\omega = n\Omega'$ is satisfied,
where $m$ and $n$ are relatively prime integers and $\omega$ and $\Omega'$
are the angular frequencies of the relative motion and tidal perturbation 
respectively.
It turns out that the tidal perturbation frequency relates to the fixed 
third-body frequency as $\Omega'=2 \Omega$, where $\Omega$ is the frequency
of the fixed third-body motion.

An averaging method that was developed for the purposes of 
studying the effect of external incident gravitational waves on
a binary system, can be applied to the equations of motion to analyze the
behavior near a resonance (Chicone, Mashhoon \& Retzloff 1997). The resultant 
averaged system 
of equations retains the `slow' variables which are left over after the system 
is averaged over a `fast' variable. The
averaged set of nonlinear equations gives rise to a solvable system, whose 
solution approximates the actual solution for sufficiently small perturbation
parameters over a certain time scale.

\section{NUMERICAL RESULTS}

One arrives at the following system of equations upon converting the
second-order Cartesian equations into a system of first-order
equations in polar coordinates:

\begin{equation}
\dot{r}=P_{r},
\end{equation}
\begin{equation}
\dot{\theta}=\frac{P_{\theta}}{r^{2}},
\end{equation}
\begin{equation}
\dot{P}_{r} = \frac{P_{\theta}^{2}}{r^{3}} - \frac{1}{r^{2}} +
\frac{1}{2}
\epsilon r [1+3\cos{(\Omega' t -2\theta)}] + \delta \frac{P_{r}}{r^{3}}
(\frac{32}{r} + 24 P_{r}^{2} + 144 \frac{P_{\theta}^{2}}{r^{2}}),
\end{equation}
\begin{equation}
\dot{P}_{\theta} = \frac{3}{2}\epsilon r^{2}\sin{( \Omega' t - 2\theta)}
+\delta \frac{P_{\theta}}{r^{3}} (\frac{24}{r} +108 P_{r}^{2} -
72\frac{P_{\theta}^{2}}{r^{2}}).
\end{equation}
The parameters $\epsilon$ and $\delta$ denote the
tidal and gravitational damping perturbation strengths, respectively.
One can in principle generate a solution to this system of differential equations 
by specifying a set of initial conditions that represent the physical state of
the system. The system of equations also includes parameters $\epsilon$, $\delta$, and $\Omega$
that characterize the strength of the perturbations and the angular frequency
of the tidal perturbation. These parameters can be picked to reflect a 
specific physical model. In fact, a simplification has been made which, in effect,
makes $\epsilon$ an independent small parameter. The original derivation
reveals that the tidal perturbation amplitude is directly proportional to
$\Omega'^{2}$, which may not be small enough in many cases.
A generalization is made in this model to accommodate the requirements
of the averaging theorem, particularly that the perturbation amplitude be 
sufficiently small. 
A series of numerical experiments can be performed in which the system evolves  
under the influence of the prescribed perturbations and initial state.
This process aims to find occurrences of resonance capture in phase space and
parameter space of the dynamical system.

To accord with the subsequent mathematical analysis, the Delaunay variable $L$ 
is graphed versus time in the numerical work. The variable $L$ is related to 
the semimajor axis $a$ of the Keplerian ellipse as $L=a^{1/2}$.
Therefore it is inversely proportional to the energy of the elliptical orbit 
as $L=(-2E)^{-1/2}$. The standard signature of the resonance capture
in terms of $L$ versus $t$ is an oscillating interval about an average value 
$L_{*}$, with an envelope that generally increases with time before 
`falling out' of 
the resonance. The average value is related to the order of the resonance. 
A sustained average orbital energy follows directly from the average $L$.

Numerical experiments are conducted with the aim of finding resonances of
different orders. In this investigation, numerical evidence of $(1:1)$, $(2:1)$, 
and $(3:1)$ resonances is found. See figures 1-3 and their captions which
show the parameters and initial conditions of the systems. It is not difficult 
to search for resonances in the parameter space of the system; in particular,
the (1:1) resonance described here is different from the one represented in a
previous paper (Wardell 2002). A cursory look at
the graphs will show that there is an increasing degree of structure,
especially with regards to the $(3:1)$ graph. One can see that the $(3:1)$
resonance contains a `dense' orbit with chaotic characteristics. That is,
the features of the graph such as the oscillations are apparently irregular
as is the amplitude. In contrast, the $(1:1)$ and $(2:1)$ orbits are much more
regular in their oscillations.

\begin{figure}

\epsfig{file=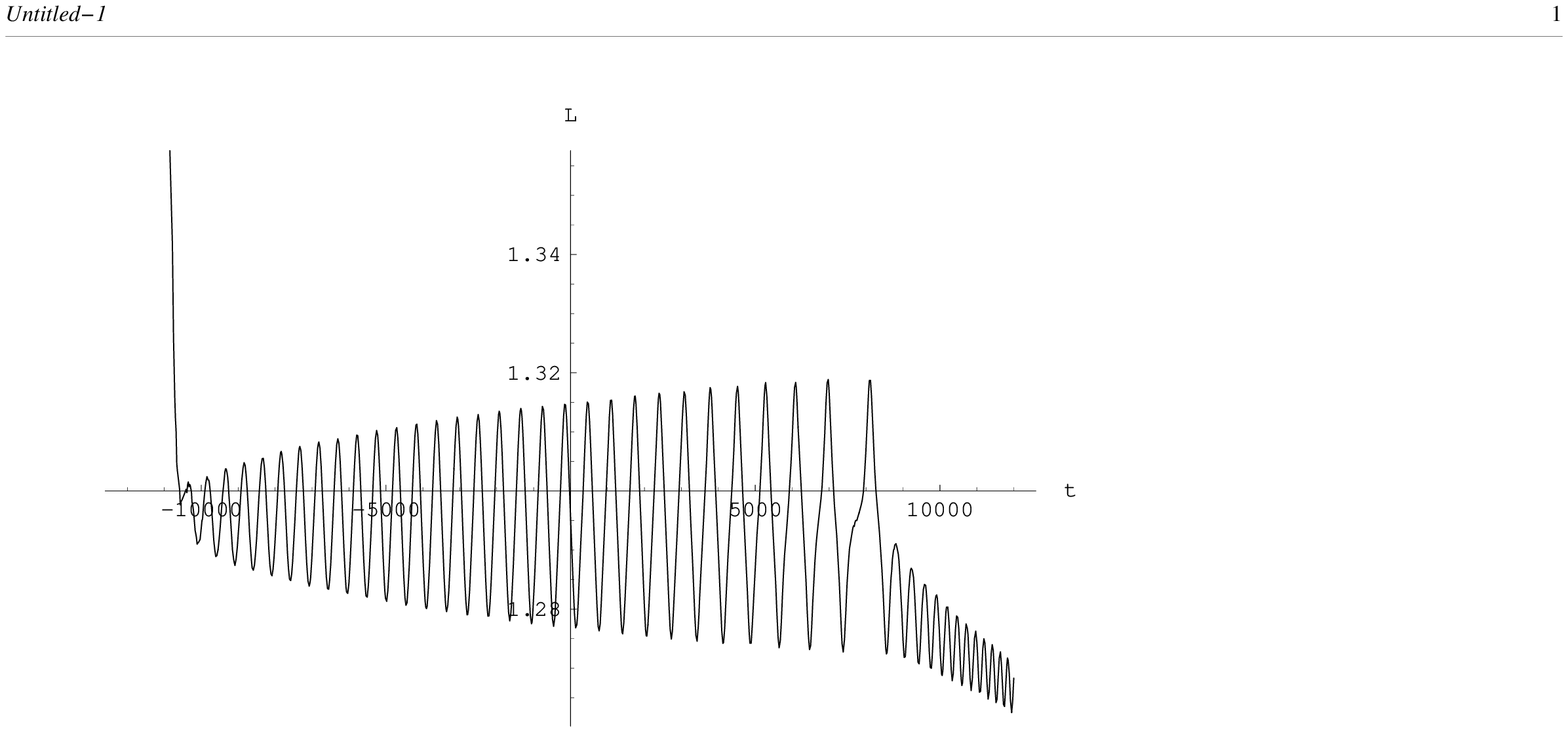,width=13cm,height=13cm,angle=-90,clip=,bbllx=50,bblly=95,
bburx=400,bbury=520}

\caption{This is a (1:1) resonance. The parameters are $\epsilon=5 \times 10^{-5}$, 
$\delta/\epsilon=10^{-3}$, and $\Omega'=0.46$. The initial conditions at $t=0$
are $(r,\theta,P_{r},P_{\theta})=(1,0.5,0.63569,1).$}

\end{figure}

\begin{figure}

\epsfig{file=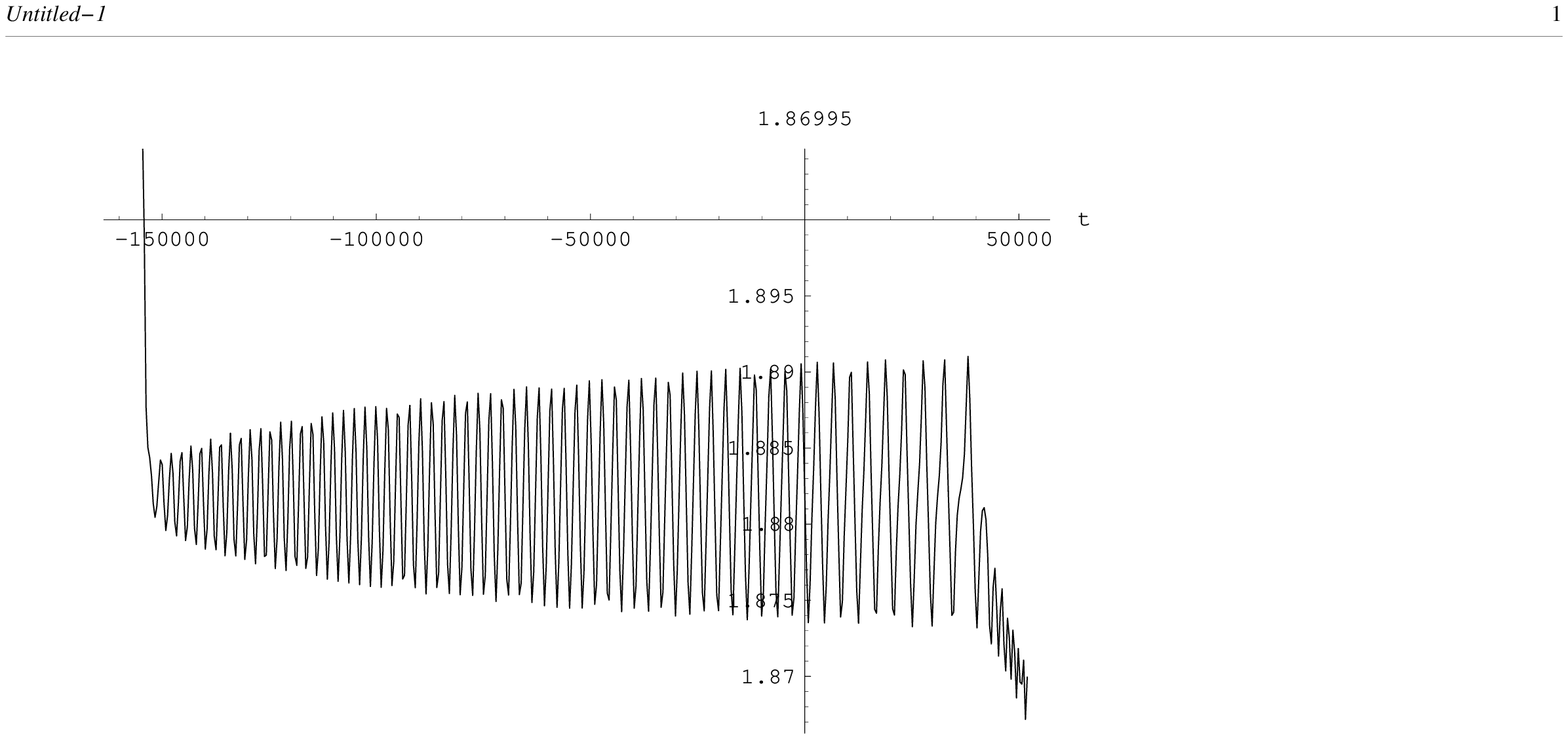,width=13cm,height=13cm,angle=-90,clip=,bbllx=50,
bblly=95,bburx=400,bbury=520}

\caption{This is a (2:1) resonance. The parameters are $\epsilon=5 \times 10^{-6}$, 
$\delta/\epsilon=10^{-3}$, and $\Omega'=0.30$. The initial conditions at $t=0$
are $(r,\theta,P_{r},P_{\theta})=(1,0.5,0.847165,1).$}

\end{figure}

\begin{figure}

\epsfig{file=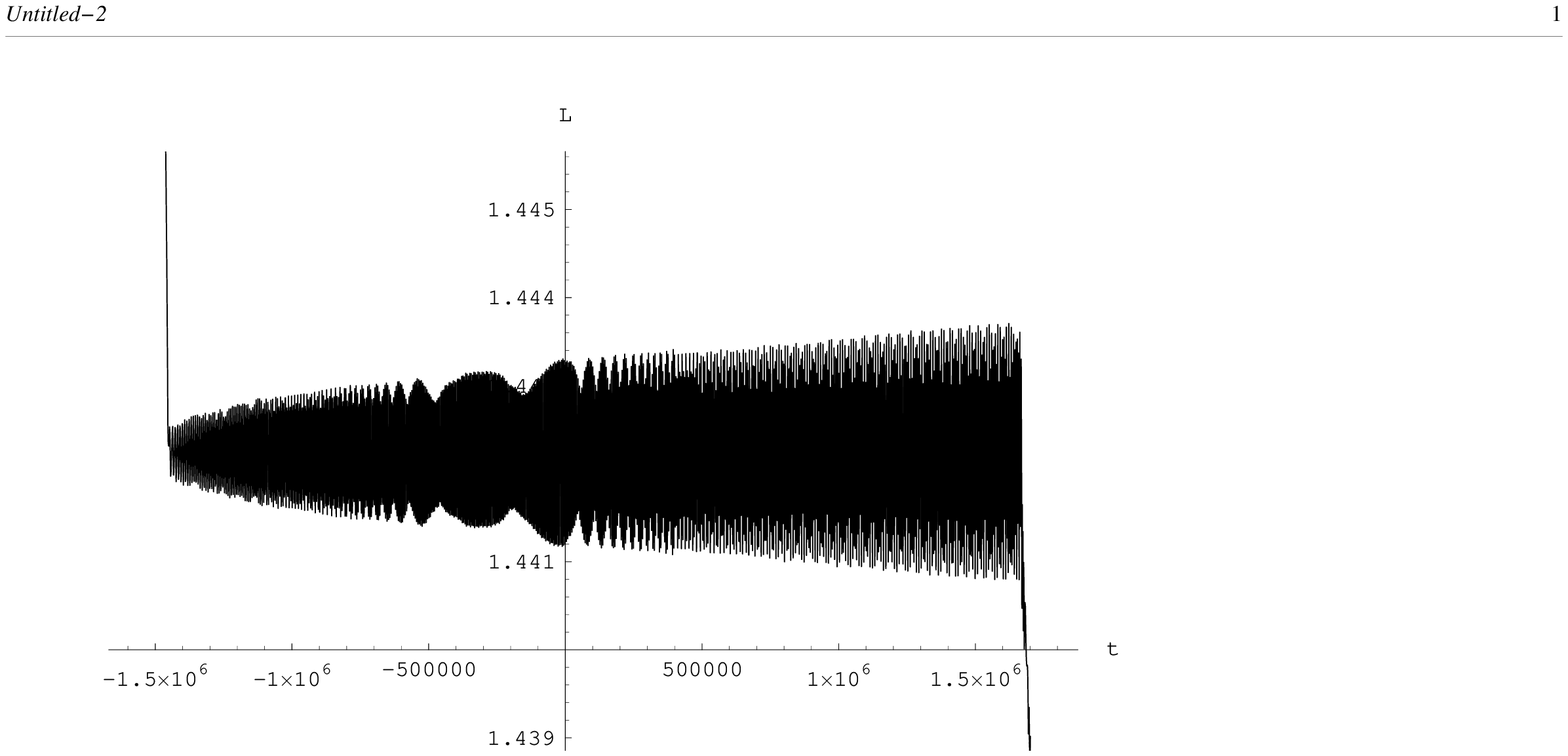,width=13cm,height=13cm,angle=-90,clip=,bbllx=50,
bblly=95,bburx=400,bbury=520}

\caption{This is a (3:1) resonance. The parameters are $\epsilon=10^{-6}$,
$\delta/\epsilon=10^{-3}$, and $\Omega'=0.10$. The initial conditions at $t=0$
are $(r,\theta,P_{r},P_{\theta})=(1,0.5,0.72059,1).$}

\end{figure}

The graphs exhibit similar qualitative behavior to which a measurement can be
applied. The oscillation during the resonance capture gives rise to an 
average frequency. Let this be $\omega_{m}$.  
Moreover, the amplitude of the \textit{envelope} of the oscillation changes 
with time and
yields another measurable quantity. Namely, the ratio between the amplitudes
of the envelope at two different times is a simple measure of the change of
the envelope that surrounds the oscillatory orbit. Let this be $R_{m}$.
The purpose of this paper is to provide analytic expressions for $\omega_{m}$
and $R_{m}$ and an approximate theoretical description of the binary orbit
while in resonance.

\section{EQUATIONS OF MOTION IN DELAUNAY VARIABLES}

If one neglects the damping terms in the equations of motion (2)-(5),
one arrives at a Hamiltonian system that can be transformed canonically into 
other coordinate systems. The coordinate transformation from polar to Delaunay 
variables is beneficial because Delaunay variables are action-angle variables
adapted to the osculating ellipse.
The Hamiltonian of the unperturbed system depends only on the actions or the
canonical momenta. This results in the momenta being constants of the motion
and the generalized coordinates being angle variables. The inclusion of 
perturbations
causes deviations from these constants --- the otherwise constant variables
evolve with time. The close relationship that the Delaunay variables have
with the elliptical elements conveniently  connect the trajectory on the 
manifold with respect to the Delaunay variables with the osculating 
ellipse description of motion that is associated with elliptical 
elements. In other words, one has at each point of time a set of coordinates
that will follow the Keplerian trajectory if the external perturbations
were removed. The orbit of the system is aptly described by an evolving
ellipse, whose descriptive coordinates change in time. For the planar system
under consideration here, the osculating ellipse can be characterized by its
semimajor axis $a$, eccentricity $e$, eccentric anomaly $u$, and true
anomaly $v$ (Danby 1988). The Delaunay elements are then defined by
$L=a^{1/2}$, $G=P_{\theta}=L(1-e^{2})^{1/2}$, $\ell=u-e\sin{u}$,
and $g=\theta-v$. Here $L$ and $G$ are action variables and the mean
anomaly $\ell$ and $g$ are the corresponding angle variables.

The equations of motion in Delaunay variables take the following general form 

\begin{equation}
\dot{L}=-\epsilon\frac{\partial \mathcal{H}_{ext}}{\partial \ell} +
\epsilon \Delta f_{L},
\end{equation}
\begin{equation}
\dot{G} = -\epsilon \frac{\partial \mathcal{H}_{ext}}{\partial g} +
\epsilon \Delta f_{G},
\end{equation}
\begin{equation}
\dot{\ell} = \frac{1}{L^{3}} + \epsilon \frac{\partial
\mathcal{H}_{ext}}{\partial L} + \epsilon \Delta f_{\ell},
\end{equation}
\begin{equation}
\dot{g} = \epsilon \frac{\partial \mathcal{H}_{ext}}{\partial G} +
\epsilon \Delta f_{g},
\end{equation}
where $\Delta=\delta/ \epsilon$. This definition is made so that one can fix 
$\Delta$
and have only one small parameter $\epsilon$ that is needed for the remaining 
analysis. The averaging theorem relies on the system having one perturbation
parameter.

The Hamiltonian associated with the binary system plus the external perturbation
due to the third mass can be expressed as
\begin{equation}
H=H_{0} + \epsilon \mathcal{H}_{ext},
\end{equation}
where the unperturbed part is
\begin{equation}
H_{0}=-\frac{1}{2L^{2}},
\end{equation}
and the tidal perturbation is
\begin{equation}
\mathcal{H}_{ext} = \mathcal{A}(L,G,\ell) + \mathcal{B}(L,G,\ell,g)\cos{\Omega t} +
\mathcal{C}(L,G,\ell,g)\sin{\Omega t}.
\end{equation}
The functions $\mathcal{A}$, $\mathcal{B}$, and $\mathcal{C}$ are given by
\begin{equation}
\mathcal{A}= 
-\frac{1}{4}L^{4}(1+\frac{3}{2}(1-\frac{G^{2}}{L^{2}})+\sum_{\nu=1}^{\infty}
\frac{\cos{\nu \ell}}{\nu^{2}} J_{\nu}(\nu e)),
\end{equation}
\begin{equation}
\mathcal{B} = -\frac{15}{8} L^{4}(1-\frac{G^{2}}{L^{2}})
 \cos{2 g}  - \frac{3}{4}L^{4} \sum_{\nu=1}^{\infty} (A_{\nu} \cos{2g}\cos{\nu \ell}
-B_{\nu}\sin{2g}\sin{\nu \ell}), 
\end{equation}
\begin{equation}
\mathcal{C}= -\frac{15}{8} L^{4}(1-\frac{G^{2}}{L^{2}})\sin{2g}
-\frac{3}{4}L^{4}\sum_{\nu=1}^{\infty}(A_{\nu}\sin{2g}\cos{\nu\ell}+
B_{\nu}\cos{2g}\sin{\nu \ell})),
\end{equation}
where
\begin{equation}
A_{\nu} = \frac{4}{\nu^{2} e^{2}}[2 \nu e(1-e^{2}) J'_{\nu}(\nu e) -
(2-e^{2}) J_{\nu}(\nu e)], 
\end{equation}

\begin{equation}
B_{\nu} = -\frac{8}{\nu^{2} e^{2}} (1-e^{2})^{\frac{1}{2}}[e J'_{\nu}
(\nu e) -\nu(1-e^{2}) J_{\nu}(\nu e)].
\end{equation}
The symbol $J_{\nu}$ represents a Bessel function of the first kind of order 
$\nu$ and $e = (1-G^{2}/L^{2})^{1/2}$ is the eccentricity.

The transformation for the general case which 
includes damping uses the following relationship which accounts for the 
non-conservative nature of the damping force: 
\begin{equation}
\frac{1}{\mu}\frac{dE}{dt} = \textbf{F} \cdot \textbf{v},
\end{equation}
where $\mu$ is the reduced mass.
One can use this relation to find how the damping terms transform with the
new coordinates. This result was derived in a previous study where the
damping terms are the same as in this case (Chicone et al. 1997). 
The terms are:

\begin{equation}
f_{L} = \frac{4}{L r^{3}}[ 1-\frac{16}{3}\frac{L^{2}}{r} +
(\frac{20}{3}L^{2} - \frac{17}{2}G^{2})\frac{L^{2}}{r^{2}} +
\frac{50}{3}\frac{L^{4}G^{2}}{r^{3}} -
\frac{25}{2}\frac{L^{4}G^{4}}{r^{4}}],
\end{equation}

\begin{equation}
f_{G}=-\frac{18 G}{L^{2} r^{3}}[1-\frac{20}{9}\frac{L^{2}}{r} +
\frac{5}{3}\frac{L^{2}G^{2}}{r^{2}}],
\end{equation}

\begin{equation}
f_{\ell} = \frac{2 \sin{v}}{e L^{3}Gr^{2}}[4 e^{2} + \frac{1}{3} (73
G^{2} - 40 L^{2})\frac{1}{r} - 2(\frac{70}{3} L^{2} -\frac{27}{2}
G^{2})\frac{G^{2}}{r^{2}} - \frac{25}{3} \frac{L^{2} G^{4}}{r^{3}} + 25
\frac{L^{2}G^{6}}{r^{4}}],
\end{equation}

\begin{equation}
f_{g}=-\frac{2 \sin{v}}{e L^{2} r^{3}} [11+(7G^{2} - \frac{80}{3}
L^{2})\frac{1}{r} - \frac{25}{3} \frac{L^{2} G^{2}}{r^{2}} + 25
\frac{L^{2} G^{4}}{r^{3}}].
\end{equation}

So one has here all of the ingredients to produce the equations of motion
in Delaunay variables. The system is formally suited for the averaging process.

\section{PARTIAL AVERAGING NEAR RESONANCE}

\subsection{FORMALISM}

When the perturbation is included, the action variables $L$ and $G$ as well
as the angle variable $g$ change in time along with the angle $\ell$. 
Their time derivatives in the equations of motion (6)-(9)
are equal to an expression of the order of the small parameter
$\epsilon$; therefore, they evolve slowly over time. The angle variable $\ell$
maintains its leading term which is of order unity --- so this describes
fast motion in relation to the other variables. 
In essence, one is interested in the evolution of the slow variables
$L$, $G$, and $g$ averaged over the fast motion of $\ell$. The motion of $\ell$ 
for the unperturbed system is just the `mean angle' for the relative orbit.
The variable $L$ is related to the energy of the orbit by $E=-1/(2L^{2})$,
$G$ is the angular momentum, and $g$ is the argument of the periastron 
of the osculating ellipse.
Averaging over the fast angle brings to light an averaged solution of
the slowly evolving variables, particularly observable variables such as
$L$.

To advance an analytical description of the modelled system, one can take
advantage of the qualitative features of the ODE --- namely, the fact that
there is one `fast' angle which evolves quickly with respect to the `slow'
variables. The `slow' angle variable can be considered to be effectively an 
action variable. This is key because to meet the conditions of the 
averaging theorem outlined below, one needs a system with only one fast angle 
variable. If one makes use of the fact that there is only one
`fast moving' angle in the system of interest, the averaging theorem can
be applied. Employing the averaging theorem, one arrives at an
averaged set of differential equations that are solvable. 
Over the correct timescale, one has an 
approximate analytical solution to the motion of the binary system of 
interest.

To illustrate how averaging is applied using the averaging theorem 
consider the following representation of a system of equations cast into 
action-angle form:

\begin{equation}
\dot{I} = \epsilon \mathcal{F} (I,\theta),
\end{equation}
\begin{equation}
\dot{\theta}=\Omega(I) + \epsilon \mathcal{G} (I,\theta),
\end{equation}
where 

\begin{equation}
I, \mathcal{F}\in \Re^{n},
\end{equation}
\begin{equation}
\theta, \Omega, \mathcal{G} \in \Re. 
\end{equation}
The symbol $\Re$ represents the set of real numbers and $\Re^{n}$ represents 
an n-dimensional real space.
The terms with the coefficient $\epsilon$ are perturbations. Furthermore, the
evolution of the action variables $I$ is slow in time with respect to
the angle variable $\theta$ that is fast moving. So one has a case where
averaging can be applied.
One can average over the angle $\theta$ and create a new differential equation
so that to first-order one has a new system in terms of $J$:
\begin{equation}
\dot{J} = \epsilon \frac{1}{2 \pi} \int_{0}^{2 \pi} \mathcal{F}
(J,\theta) d\theta.
\end{equation}
The new averaged variable $J$ coincides with the actual variable $I$ at
$t=0$:
\begin{equation}
J(0) = I(0) = I_{0}.
\end{equation}
The Averaging Theorem states that one can create an averaged system whose 
solution will closely follow the actual system's solution.
That is the trajectory of the new averaged variable $J(t)$ will stay close to
the true trajectory $I(t)$ within a distance of order $\epsilon$. This will
be valid within a timescale of $\epsilon^{-1}$.
So the following inequalities describe the averaged system where $C$ and $C'$
are constants:
\begin{equation}
|I(t) - J(t)| \leq C\epsilon,
\end{equation}
\begin{equation}
0 \leq t \leq \frac{C'}{\epsilon}.
\end{equation}

Inspection of the differential equation in the Delaunay variables reveals
that the `fast' variable is $\ell$ while the others are `slow'.
It is of interest to study the behavior of the system's motion near a resonance.
At a resonance, as can be seen in the numerical case, the value for $L$ stays
fixed on average.  
For a fixed $L_{*}$ associated with a particular resonance, one arrives at
the following resonance condition in terms of $L_{*}$:
\begin{equation}
m \frac{1}{L_{*}^{3}} = n \Omega'
\end{equation}
where $\omega=L^{-3}$.
The unperturbed equation for $\ell$ yields the solution
\begin{equation}
\ell(t)=\frac{1}{L^{3}}t + \ell_{o},
\end{equation}
where $\ell_{o}$ is an integration constant.

To highlight the motion near resonance, one can consider the deviations that
occur near the resonance. The averaging that one performs at resonance, 
is called partial averaging.
To investigate the behavior of the binary near the resonance manifold 
$(L_{*},G,\ell,g)$ one makes the transformation
\begin{equation}
L = L_{*} + \epsilon^{1/2} D, \; \;
\ell = \frac{1}{L_{*}^{3}} t + \varphi.
\end{equation}
This introduces the variable $D$ which represents the `deviation' from the
resonance value of $L_{*}$ and the variable $\varphi$ which is related to
a deviation from the unperturbed mean angle $\ell$.
The transformation involves $\epsilon^{1/2}$ before the $D$ variable and
unity before the $\varphi$ variable so that the transformed ODE will
contain the differential equations in $D$ and $\varphi$ with leading terms
to the same order. The leading order becomes the small parameter for the
averaged system, that is $\epsilon^{1/2}$. This gives the proper form for
averaging.

The derivation involves an expansion around $L_{*}$, therefore it is possible to
mathematically generate any number of terms in the expansion. Since the
model equations that reflect the physics under investigation are given to highest
order $\epsilon$, one does not benefit from a physics point of view to expand
beyond $\epsilon$ in the transformation. 

After the transformation is made one obtains the following expression for the 
transformed system at resonance (Chicone et al., 1997): 

\begin{equation}
\dot{D} = -\epsilon^{1/2} F_{11} -\epsilon D F_{12} +
O(\epsilon^{3/2}),
\end{equation}
\begin{equation}
\dot{G} = -\epsilon F_{22} + O(\epsilon^{3/2}),
\end{equation}
\begin{equation}
\dot{\varphi} = -\epsilon^{1/2} \frac{3}{L_{*}^{4}} D +
\epsilon(\frac{6}{L_{*}^{5}} D^{2} + F_{32}) + O(\epsilon^{3/2}),
\end{equation}
\begin{equation}
\dot{g} = \epsilon F_{42} + O(\epsilon^{3/2}).
\end{equation}
Here $F_{ij}$ are defined by:

\begin{equation}
F_{11} := \frac{\partial \mathcal{H}_{ext}}{\partial
\ell}(L_{*},G,\frac{n\Omega'}{m}t + \varphi,g,t) - \Delta
f_{L}(L_{*},G,\frac{n\Omega'}{m} t + \varphi,g),
\end{equation}

\begin{equation}
F_{12}:= \frac{\partial F_{11}}{\partial L}(L_{*},G,\frac{n\Omega'}{m}t +
\varphi,g,t),
\end{equation}

\begin{equation}
F_{22}:= \frac{\partial \mathcal{H}_{ext}}{\partial
g}(L_{*},G,\frac{n\Omega'}{m}t+\varphi,g,t) - \Delta
f_{G}(L_{*},G,\frac{n\Omega'}{m}t +\varphi,g),
\end{equation}

\begin{equation}
F_{32}:=\frac{\partial \mathcal{H}_{ext}}{\partial
L}(L_{*},G,\frac{n\Omega'}{m}t + \varphi,g,t) + \Delta
f_{\ell}(L_{*},G,\frac{n\Omega'}{m}t + \varphi,g),
\end{equation}

\begin{equation}
F_{42}:=\frac{\partial \mathcal{H}_{ext}}{\partial
G}(L_{*},G,\frac{n\Omega'}{m}t+\varphi,g,t) + \Delta
f_{g}(L_{*},G,\frac{n\Omega'}{m}t + \varphi,g).
\end{equation}

The next step is to derive the actual second order partially averaged
equations for the system. To do this, one can make use of a special coordinate 
change
known as an averaging transformation. The purpose of such a transformation
is to render the system by way of a coordinate change into a form that is
averaged to first order. Since averaging to second order is desired here,
one needs to average the entire transformed system and drop terms of order
higher than second. 

To apply the averaging transformation one must make the following
definitions:

\begin{equation}
\overline{F}_{ij}:=\frac{\Omega'}{2\pi m}\int_{0}^{\frac{2 \pi 
m}{\Omega'}}
F_{ij}(G,\frac{n \Omega'}{m} s + \varphi,g,s)ds,
\end{equation}

\begin{equation}
\lambda(G,\varphi,g,t):= F_{11}(G,\frac{n \Omega'}{m}t + \varphi,g,t)-
\overline{F}_{11},
\end{equation}

\begin{equation}
\Lambda(G,\varphi,g,t):=\int \lambda(G,\varphi,g,s)ds,
\end{equation}
such that
\begin{equation}
\int_{0}^{\frac{2\pi m}{\Omega'}} \Lambda(G,\varphi,g,s)ds = 0.
\end{equation}
That is, the average of $\Lambda$ vanishes.
The averaging transformation becomes:
\begin{equation}
D=\hat{D} - \epsilon^{1/2} \Lambda(\hat{G},\hat{\varphi},\hat{g},t), \; \; \;
\;
G=\hat{G}, \; \; \; \;
\varphi=\hat{\varphi}, \; \; \; \;
g=\hat{g}.
\end{equation}
This transformation has the property that its average becomes the identity
transformation.
The averaging transformation changes equations (23)-(26) into the following:
\begin{equation}
\dot{\hat{D}}=-\epsilon^{1/2} \overline{F}_{11}-\epsilon \hat{D}(F_{12}+
\frac{3}{L_{*}^{4}}\frac{\partial \Lambda}{\partial \varphi}) 
+O(\epsilon^{3/2}),
\end{equation}

\begin{equation}
\dot{\hat{G}}=-\epsilon F_{22} + O(\epsilon^{3/2}),
\end{equation}

\begin{equation}
\dot{\hat{\varphi}}=-\epsilon^{1/2} \frac{3}{L_{*}^{4}}\hat{D} +\epsilon
(\frac{6}{L_{*}^{5}}\hat{D}^{2} + F_{32} + \frac{3}{L_{*}^{4}} \Lambda)
+O(\epsilon^{3/2}),
\end{equation}

\begin{equation}
\dot{\hat{g}}=\epsilon F_{42} + O(\epsilon^{3/2}).
\end{equation}

Noting as well that the averages of $\lambda$ and $\partial \Lambda /
\partial \varphi$ vanish by construction of the transformation, one can
average the new system, drop terms of order $\epsilon^{3/2}$ and higher, and
arrive at the second order partially averaged system: 

\begin{equation}
\dot{D} = -\epsilon^{1/2} \bar{F}_{11} - \epsilon D
\bar{F}_{12},
\end{equation}

\begin{equation}
\dot{G}=-\epsilon \bar{F}_{22},
\end{equation}

\begin{equation}
\dot{\varphi}= -\epsilon^{1/2} \frac{3}{L_{*}^{4}}D +
\epsilon(\frac{6}{L_{*}^{5}} D^{2} + \bar{F}_{32}),
\end{equation}

\begin{equation}
\dot{g}= \epsilon \bar{F}_{42}.
\end{equation}

\subsection{COMPUTATION OF AVERAGED EQUATIONS}

Computation of the averages of $F_{ij}$ according to equation (43) results in 
the second order partially averaged system:

\[
\dot{D}=-\epsilon^{1/2}[\frac{3m}{8} L_{*}^{4}(A_{m}+B_{m})\sin{(2g+m
\varphi)} +
\frac{\Delta}{G^{7}}(8+\frac{73}{3}e^{2}+\frac{37}{12}e^{4})]
\]

\begin{equation}
-\epsilon D\{ \frac{3m}{8} \frac{\partial}{\partial
L}[L^{4}(A_{m}+B_{m})]|_{L_{*}}\sin{(2g+m \varphi)} + \frac{\Delta}{3
L_{*}^{3}G^{5}}(146+37 e^{2})\},
\end{equation}

\begin{equation}
\dot{G}=-\epsilon[\frac{3}{4}L_{*}^{4}(A_{m}+B_{m})\sin{(2g+m
\varphi)}
+ \frac{\Delta}{L_{*}^{3}G^{4}}(8+7e^{2})],
\end{equation}

\[
\dot{\varphi}=-\epsilon^{1/2} \frac{3}{L_{*}^{4}}D +
\epsilon\{\frac{6}{L_{*}^{5}}D^{2} - \frac{5}{2}L_{*}^{3} +
\frac{3}{4}L_{*}G^{2} \]
\begin{equation}
-\frac{3}{8} \frac{\partial}{\partial
L}[L^{4}(A_{m}+B_{m})]|_{L_{*}} \cos{(2g+m\varphi)}\},
\end{equation}

\begin{equation}
\dot{g}=\epsilon[\frac{3}{4} G L_{*}^{2} -
\frac{3}{8}L_{*}^{4}\frac{\partial}{\partial G}(A_{m}+B_{m}) \cos{(2g+m
\varphi)}],
\end{equation}
where $e=(1-G^{2}/L_{*}^{2})^{1/2}$. Recall that $A_{\nu}$ and
$B_{\nu}$ are functions of the eccentricity $e$.
At resonance the value of $L$ is $L_{*}$ and the type of resonance is
governed by the resonance relation $mL_{*}^{-3}=n\Omega'$. Through the
integration to determine the averages of $F_{ij}$ in equation (43), it is found 
\textit{that the only resonances that contribute in the averaging process are 
the $(m:1)$ resonances}.

One can make a simplification by noticing that the arguments of the
trigonometric functions all combine as $2g+m \varphi$. One can then
define a variable $\Phi$ such that $\Phi = 2g+m \varphi$. If one
multiplies the differential equations for $\dot{g}$ and $\dot{\varphi}$
by the appropriate factor
and adds them, one has a reduced system of differential equations that
consists of three equations. One of the variables is now an angle $\Phi$.
The system reduces to

\[\dot{D}=-\epsilon^{1/2}[\frac{3m}{8} L_{*}^{4}(A_{m}+B_{m}) \sin{\Phi} +
\Delta \Gamma_{\alpha}]\]
\begin{equation}
-\epsilon D\{\frac{3m}{8} \frac{\partial}{\partial
L}[L^{4}(A_{m}+B_{m})]|_{L_{*}}\sin{\Phi} + \Delta \Gamma_{\beta}\},
\end{equation}

\begin{equation}
\dot{G}=-\epsilon[\frac{3}{4} L_{*}^{4}(A_{m}+B_{m})\sin{\Phi} + \Delta
\Gamma_{G}],
\end{equation}

\[\dot{\Phi}= -\epsilon^{1/2} \frac{3m}{L_{*}^{4}}D
+\epsilon\{\frac{6m}{L_{*}^{5}}D^{2}- \frac{5m}{2} L_{*}^{3}
+\frac{3m}{4} L_{*} G^{2} + \frac{3}{2}GL_{*}^{2}\]
\begin{equation}
-\frac{3m}{8} \frac{\partial}{\partial
L}[L^{4}(A_{m}+B_{m})]|_{L_{*}}\cos{\Phi} -
\frac{3}{4} L_{*}^{4} \frac{\partial}{\partial G}(A_{m}+B_{m})\cos{\Phi}\},
\end{equation}
where the gravitational damping terms are:
\begin{equation}
\Gamma_{\alpha} = \frac{1}{G^{7}}(8+\frac{73}{3} e^{2} + \frac{37}{12}
e^{4}),
\end{equation}

\begin{equation}
\Gamma_{\beta} = \frac{1}{3L_{*}^{3}G^{5}}(146+37e^{2}),
\end{equation}

\begin{equation}
\Gamma_{G} = \frac{1}{L_{*}^{3}G^{4}}(8+7e^{2}).
\end{equation}

The result of the coordinate transformations and averaging is a system of 
three differential equations which describe the averaged motion to second 
order. The resultant set of ODEs represents the evolution of these `slow' 
variables.

\section{THE AVERAGED SYSTEM}

The averaged system derived in the previous section is in the form of an
expansion in the powers of the small parameter $\epsilon^{1/2}$ 
to second order. One can investigate the
two different orders of the partial averaging that are calculated here,
and arrive at results that can be compared to those found numerically.
The different orders bring to light different features of the solution.

\subsection{FIRST-ORDER SYSTEM}

To extract the first-order partially averaged system one needs to only 
keep terms of order $\epsilon^{1/2}$, the small parameter of the
perturbation expansion around the resonance manifold.
One obtains the following system:

\begin{equation}
\dot{D}=-\epsilon^{1/2}[\frac{3m}{8} L_{*}^{4}(A_{m}+B_{m})\sin{\Phi} +
\Delta \Gamma_{\alpha}],
\end{equation}

\begin{equation}
\dot{G} = 0,
\end{equation}

\begin{equation}
\dot{\Phi}=-\epsilon^{1/2} \frac{3m}{L_{*}^{4}}D,
\end{equation}
where $A_{m}$ and $B_{m}$ follow from equations (16) and (17) and
$\Gamma_{\alpha}$ is defined in equation (63).

One immediate consequence of this system, as can be seen in (67), is that 
the orbital angular momentum $G$ is constant at resonance; hence, the
eccentricity of the orbit is constant as well.
As a result, one can rewrite the system in an equivalent way in terms of 
two canonically
conjugate variables $D$ and $\Phi$. The corresponding Hamiltonian is:
\begin{equation}
H= \frac{1}{2}D^{2} + \epsilon \lambda \cos{\Phi} - \epsilon \tau \Phi,
\end{equation}
where $\lambda$ and $\tau$ are constants
\begin{equation}
\lambda = \frac{9m^{2}}{8}(A_{m}+B_{m}),
\end{equation}
\begin{equation}
\tau= \frac{3m}{L_{*}^{4}} \Delta \Gamma_{\alpha}. 
\end{equation}
The canonical equations of the equivalent system are:
\begin{equation}
\dot{D}=\epsilon \lambda \sin{\Phi} +\epsilon \tau,
\end{equation}
\begin{equation}
\dot{\Phi}=D.
\end{equation}
Two pendulum-like equations of motion result, one in
$D$ and one in $\Phi$. The one in $D$ traces the orbit of the deviation from
the resonant value $L_{*}$ as defined in (33) and is expressed as:
\begin{equation}
\ddot{D}+\epsilon \omega_{D}^{2} D=0,
\end{equation}
where
\begin{equation}
\omega_{D}^{2}=-\lambda \cos{\Phi}.
\end{equation}
The differential equation in $\Phi$ describes a pendulum with constant torque.
Of course, to have oscillatory behavior in both cases, the condition 
$\omega_{D}^{2}>0$ must be met.
The first-order approximation of the resonance behavior in $D$ consists 
of an oscillation with slowly varying frequency and constant amplitude.
A comparison between the analytical formula, particularly that of the frequency
(75),  and the corresponding numerical result can 
be made to test the quantitative agreement between the two methods.

\subsection{SECOND-ORDER SYSTEM}

When one considers the second-order averaged equations, one can derive a
formula for the damped or antidamped oscillator. It goes as follows:

\begin{equation}
\ddot{D} + \epsilon \gamma \dot{D} + \epsilon \omega_{D}^{2} D =0.
\end{equation}
There is also a corresponding equation for $\Phi$.
The frequency $\omega_{D}$ is the same as in the previous case, 
but the coefficient to the damping term can be found from
\begin{equation}
\gamma = \frac{3}{8}m \frac{\partial}{\partial L}\left[L^{4}
(A_{m}+B_{m})\right]_{L_{*}}\sin{\Phi} + \Delta
\frac{1}{3G^{5}L_{*}^{3}}(146+37e^{2}).
\end{equation}
This describes an oscillation whose amplitude changes with time due primarily
to damping or antidamping.
Also, $G$ is no longer constant in the second order as shown in 
equation (61). While the orbit is in resonance, its orbital angular momentum
and eccentricity change slowly over time.

For the purpose of finding an approximate solution to the differential 
equation (76) one can assume that $\gamma$ and $\omega_{D}$ are constant. 
That solution is:
\begin{equation}
D(t) = \mathcal{A}e^{-\frac{1}{2}\epsilon \gamma t} \cos{(\epsilon^{1/2} 
\omega_{D} t + constant)}.
\end{equation}
 The attention will
be focused here on $D$ because the numerical data to be compared with
the averaged results involve the deviations from the resonance value, 
which is what $D$ measures.
From this equation, one can see that there is a sinusoidal oscillation
multiplied by a time-dependent exponential amplitude. The character of this 
function depends on the magnitude and sign of $\gamma$ given in equation (77).
The amplitude increases if $\gamma < 0$ and decreases if $\gamma >0$.
The magnitude of $\gamma$ affects the rate at which the amplitude changes.
More generally, $\gamma$ itself is a function of time; this may result in
local variations of the sign of $\gamma$.

\section{NUMERICAL MEASUREMENTS}

The measured frequency $\omega_{m}$ is 
simply an average frequency determined over several periods of the numerical 
$L$ versus
$t$ graph. One must keep in mind that the averaging is valid over a time interval
of $\epsilon^{-1/2}$. 
To measure the ratio between the value of the envelope of $D$ at two points, 
that is $R_{m}$, 
one simply measures the respective values at different times and divides them. 

How does one compare the numerical results to the analytical? First of all,
it is necessary to recall the derivation of the averaged equations. They
are derived under the assumption that the orbit is in resonance. 
Since $L=L_{*}$ determines the resonance manifold, the points $(L,G,\ell,g)$ 
such that $L=L_{*}$ are on the resonance manifold. 
In the $L$ versus $t$ graph the orbit passes through the resonance value
$L_{*}$ as it oscillates. It is these points of the numerical
solution that are on the resonance manifold that apply to the analytical
formulae for $\omega_{D}$ and $\gamma$.

Recall that the first order averaged equations give a formula for the
frequency of the solution. It is $\omega_{D}^{2}=-\epsilon \frac{9}{4}K_{m}(e)
\cos{\Phi}$. This formula depends on $e$ and $\Phi$ at the resonance
manifold. To produce a value of $\omega_{D}$ one needs the values of
$e$ and $\Phi$ on the resonance manifold. In essence, what is being
investigated is the agreement between the measured frequency of a resonance
recorded in a graph and a prediction made by a derived formula. The derived
formula is based on the orbit in resonance. For the purposes of testing this
agreement one can take the values of $(L,G,\ell,g)$ that occur at resonance
in the numerical case and apply them to the derived formula.

The solution to the differential equation derived from the second-order case
gives rise to a formula that describes the behavior of the envelope of the
resonance orbit.
To find the analytical formula for this ratio let $\mathcal{V}(t)$ be a function 
that describes the envelope of $D(t)$. That is, $\mathcal{V}(t) = 
\mathcal{A} \exp(-\frac{1}{2}\epsilon \gamma t)$.
One can compute the ratio between $\mathcal{V}(t_{1})$ and 
$\mathcal{V}(t_{2})$, 
where $\Delta t = t_{2}-t_{1}$, by dividing the two expressions for 
$\mathcal{V}(t_{1})$ and $\mathcal{V}(t_{2})$.
One comes up with $R_{D}=\exp(-\frac{1}{2} \epsilon \gamma \Delta t)$.
The determination of $\gamma$ calls for the numerical resonance manifold 
values of $e$, $G$, and $\Phi$.

The examples that were computed for this study in figures 1-3 yielded the 
following results where $\omega_{m}$ is the angular frequency and $R_{m}$ is the
ratio of the amplitudes of the envelope at two different times for the $(m:1)$
resonance. Also, the following 
definitions are used:
\[\frac{\Delta \omega}{\omega} = \frac{| \omega_{D} - \omega_{m}|}{\omega_{D}},\]
\[\frac{\Delta R}{R} = \frac{|R_{D}-R_{m}|}{R_{D}} .\]

The results are:

\[
\begin{array}{crr} Resonance & \Delta \omega / \omega & \Delta R/R \\
                       (1:1) & 0.075 & 0.088 \\
                   (2:1)  & 0.0002 & 0.0186 \\
                   (3:1)  & 0.57 & ---
\end{array}
\]

A condition for the averaging theorem to be valid is that $\epsilon$ be
sufficiently small. Therefore, when computing the numerical results in search
of resonances, one can pose the question of whether the numerical choice of
$\epsilon$ is small enough. How small is sufficiently small? By lowering the
value of $\epsilon$, which in turn leads to a more cumbersome and 
time-consuming numerical calculation, one can most likely improve the
numerical results. As can be seen in the summary of results, the (1:1) and
(2:1) examples show good agreement between the numerical and analytical results.
The (3:1) case, however, deviates quite far from agreement. It is known from
numerical evidence that the higher-order resonances yield more elaborate 
structure and are more likely to exhibit chaos than low-lying resonances
(Chicone et al.,1997). In this case, the numerical results could be precarious in their
agreement with analytical predictions. Perhaps the choice for $\epsilon$ was
not sufficiently small to meet the conditions of the averaging theorem.

\section{ANALYSIS OF RESONANCE CAPTURE}

Librational motion on the phase plane of the first-order model is indicative
of an orbit captured into resonance (Chicone et al., 1997). The pendulum-with-torque system
shown in (72) and (73) must have an elliptical fixed point to have these
librations. Furthermore, a hyperbolic fixed point is expected with a homoclinic
orbit to give rise to an area in the phase space where orbits pass through
resonance. 
A fixed point on the phase plane is given by $(D,\Phi) = (0,\Phi_{0})$, where
$\Phi_{0}$ satisfies:
\begin{equation}
\sin{\Phi_{0}}=-\tau / \lambda.
\end{equation}

In general, three different scenarios can arise as the result of the pendulum
system.
For there to be a solution for $\Phi_{0}$ the inequality 
$|\tau/\lambda| \leq 1$ must hold.
For the cases where $|\tau / \lambda| < 1$ or 
$|\tau/\lambda| = 1$, there are fixed points. 
Consider the case where the inequality $|\tau/\lambda| < 1$
holds. The phase cylinder for this shows that 
there exist orbits that are oscillatory and stay in the
resonance capture region, and also orbits that pass through the resonance 
(see Figure 4).
This shows that whether the orbit will be captured into resonance 
depends on the initial conditions.
\begin{figure}

\epsfig{file=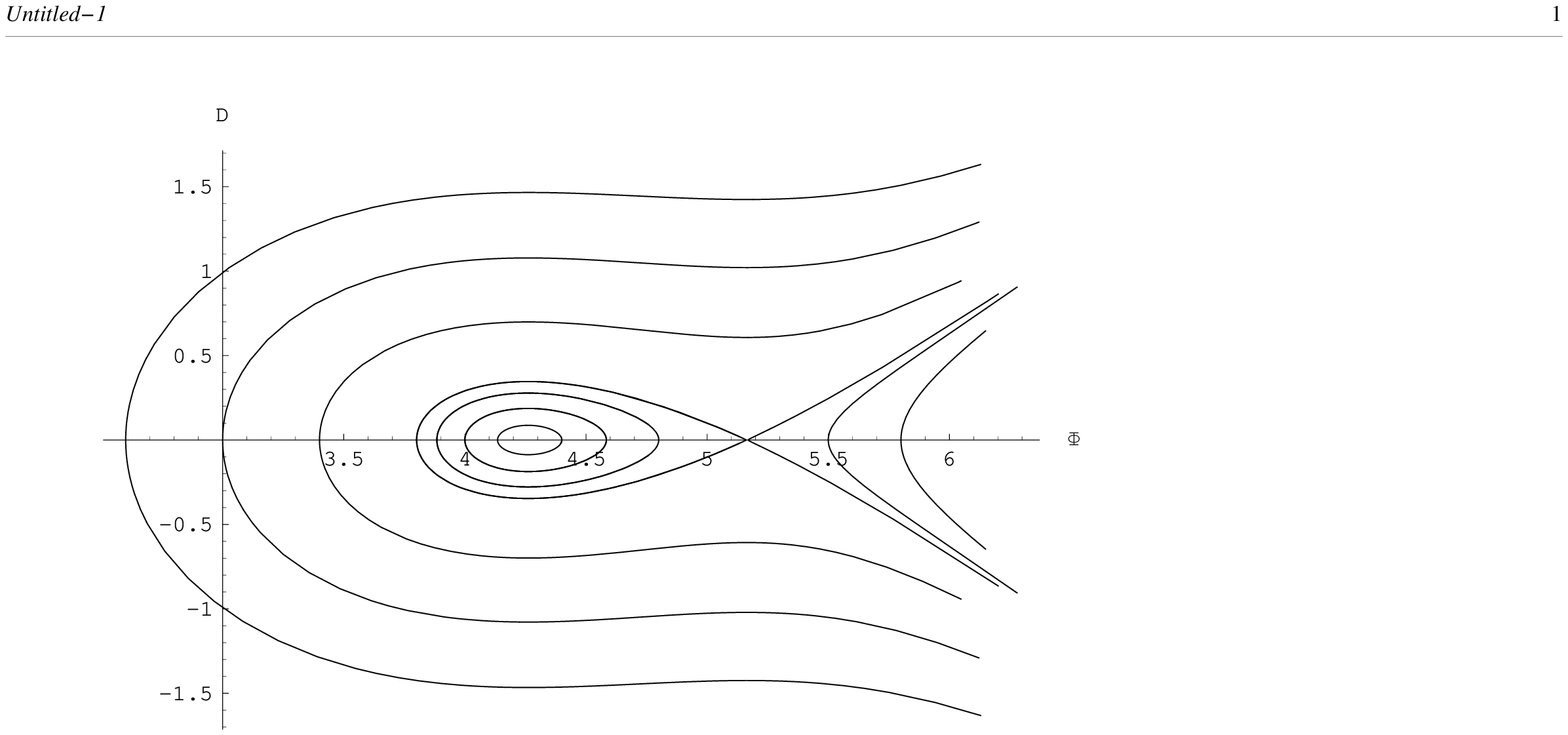,width=8cm,height=8cm,angle=-90,clip=,bbllx=50,bblly=95,
bburx=400,bbury=500}

\caption{This is an example of a phase cylinder representing a pendulum
equation with constant external torque. Specifically, the system is
$\dot{D}=a \sin{\Phi}+b$ and $\dot{\Phi}=D$, where $a=1$ and $b=0.9$.
In this example $0$ and $2 \pi$ are identified on the horizontal $\Phi$ axis.
The vertical axis is $D=\dot{\Phi}$. The elliptical orbits represent 
periodic solutions around a rest point. There is a seperatrix beyond which the 
solutions are not periodic.}

\end{figure}
The case such that $|\tau/\lambda| > 1$ can arise where there is no 
solution to the equation and therefore no fixed point. 

The first-order system of equations that gives rise to these pendulum-like
equations displays in an approximate way the physics of the original
astrophysical model. From the first-order system one can see from equation
(71) that the torque $\tau$ is linearly related to the gravitational 
radiation damping by the coefficient $\Delta$. If there is sufficiently strong 
damping, and hence a large $\Delta$, there will be no resonance
capture because $\tau / \lambda$ can in all cases exceed unity. All 
solutions will pass through the resonances and the orbit will follow a 
path such that the semimajor axis will shrink to zero. This would be the 
expected result if gravitational radiation damping were the sole 
perturbation. Recall also that the binary system will be captured only 
into the $(m:1)$ resonances according to the averaged equations. 

The second-order averaged 
system includes antidamping effects as well as a slowly drifting $\lambda$ and
$\tau$ (Chicone et al. 2000). It offers a description 
of how the actual system can fall out of resonance by leaving the 
capture region by its anti-damped motion. The full non-linear 
system
exhibits effects like resonance capture and exit from resonance, dynamics
whose full description elude current analytical techniques. However, the 
second-order system 
explains with some success the system's behavior in resonance, particularly 
the frequency and the rate of antidamping.

There are interesting features in the $(3:1)$ resonance, particularly its
alternating sets of grouped oscillations in an `excited' phase and
`relaxed' phase.
The slowly changing variables of $L$, $G$, and $g$, and the fast variable
$\ell$ as seen in the Delaunay form of the system (6)-(9)
make it a candidate for the phenomenon of bursting. One
may speculate from the characteristic features in figure 3 that bursting is
present in this $(3:1)$ resonance (Izhikevich 2001).
 
The phenomenon of resonance capture describes the system's state where the
dissipative influence of gravitational radiation damping is offset on the
average by the tidal perturbation of the orbiting third mass; that is,
the theory of capture into resonance 
describes how the orbit's loss of energy via gravitational radiation
reaction is countered by a deposit of energy from the orbit of the third mass.

\section{CONCLUSION}

The averaging method produces a set of ordinary differential equations that
describe the behavior of the system while in resonance. The quantity $D$ 
tracks the oscillatory motion of $L$ about $L_{*}$.
The level of approximation given by the first-order averaged equations
highlights the oscillatory characteristics of the actual solution whereas
the second-order averaged equations reveals the slow changes in amplitude. 
The numerical
solution shows more structure and detail that is inherent in the actual
solution. There are prominent features of the actual solution that one can 
compare qualitatively and quantitatively to that of the averaged system. 
Though the first-order
solution only predicts oscillations of constant amplitude and 
slowly changing frequency
and the second-order solution predicts an exponential trend of damping or
antidamping, they correspond on the appropriate timescale to the most 
prominent 
features of the numerical solution. This makes sense in the light of what the
averaging method intends to do --- capture the `slow' motion.

Resonance capture and chaos are two signature results that arise from a
system in resonance. Perhaps the most noted example of this phenomenon is
the damped driven oscillator. The astrophysical system that is modelled
in this study by a damped and driven Kepler system displays similar
characteristic effects --- namely resonance capture. Under the
non-Hamiltonian perturbation, the nonlinear Hamiltonian system can exhibit
both resonance capture and chaos (Chicone et al. 2000). This can be manifest in the attempt
to compare the numerical case with the analytical. An orbit near a higher-order 
resonance such as $(3:1)$ will more likely be chaotic and therefore prone to
bear `erroneus' results when compared with the partially averaged equations. 
The dense structure shown in figure 3 that corresponds
to $(3:1)$ suggests such chaotic motion. However, good results followed from
the analysis of the $(1:1)$ and $(2:1)$ resonances. 

The averaged system to first order shows that at each $(m:1)$ resonance one sees
motion like a pendulum with constant torque. To the second order, antidamping
and nonlinearity may cause disruption to the resonance.
The full nonlinear system includes more effects that are not seen in the
averaged systems which may contribute to the disruption of the resonance.
The analytical formulas for the first and second-order averaged systems outline
structure that is part of the actual solution. It is also of interest to
determine how well the
results from the analytical part agree with the numerical. The complexity of
the actual nonlinear equations and their solution makes analytical formulae
helpful and illuminating. Both analytical and numerical techniques given here
present a consistent description of the nonlinear effects associated with
resonance capture.

\vspace{.2cm}

\begin{flushleft}
\textbf{\Large{ACKNOWLEDGEMENTS}}\\

I would like to thank B. Mashhoon and C. Chicone for their helpful and 
stimulating discussions and critical readings of this manuscript.

\vspace{.4cm}

\textbf{\Large{REFERENCES}}\\

Chicone C.,Mashhoon B.,Retzloff D.G., 1997, Class. Quantum Grav., 
\\ \hspace{.5cm}14, 1831\\
Chicone C.,Mashhoon B.,Retzloff D.G., 2000, J. Phys. A: Math. Gen., 
\\ \hspace{.5cm}33, 513\\
Chicone C.,Kopeikin S.,Mashhoon B.,Retzloff D.G., 2001, Phys. Lett. A,
\\ \hspace{.5cm} 285,17\\
Danby J.M.A. 1988, Fundamentals of Celestial Mechanics, 2nd ed., 
\\ \hspace{.5cm} Willmann-Bell, Richmond\\
Izhikevich E.M., 2001, SIAM Rev., 43, 315\\
Wardell Z., 2002, MNRAS, in press

\end{flushleft}

\end{document}